\documentclass[aps,prb,twocolumn,showpacs,showkeys,fleqn,floatfix]{revtex4-1}

\usepackage[dvips]{epsfig}
\usepackage[dvips]{color}
\usepackage{amsmath}
\usepackage{multirow}
\usepackage{nicefrac}

\newcommand{\figwidth}{0.98\columnwidth}
\newcommand{\bacusio}{BaCu$_2$Si$_2$O$_7$}
\newcommand{\vect}[1]{\mathbf{#1}}
\usepackage{mathptmx}
\usepackage{helvet}%[scaled=.90]{helvet}
\usepackage{courier}

\begin{document}
\title{Phase diagram of low-dimensional antiferromagnets with competing order parameters:
A Ginzburg-Landau-theory approach}

\author{V.~N.~Glazkov}
\email{glazkov@kapitza.ras.ru}
\affiliation{Kapitza Institute for Physical Problems RAS, Kosygin
str.\ 2, 119334 Moscow, Russia}
\author{F.~Casola}
\affiliation{Laboratorium f\"ur Festk\"orperphysik, ETH
H\"onggerberg, CH-8093 Z\"urich, Switzerland}
\affiliation{Paul Scherrer Institut, CH-5232 Villigen PSI, Switzerland}%
\author{H.-R.~Ott}
\affiliation{Laboratorium f\"ur Festk\"orperphysik, ETH
H\"onggerberg, CH-8093 Z\"urich, Switzerland}
\affiliation{Paul Scherrer Institut, CH-5232 Villigen PSI, Switzerland}%
\author{T.~Shiroka}
\affiliation{Laboratorium f\"ur Festk\"orperphysik, ETH
H\"onggerberg, CH-8093 Z\"urich, Switzerland}
\affiliation{Paul Scherrer Institut, CH-5232 Villigen PSI, Switzerland}%

\date{\today}

\begin{abstract}
We present a detailed analysis of the phase diagram of
antiferromagnets with competing exchange-driven and field-induced
order parameters. By using the quasi-1D antiferromagnet \bacusio{}
as a test case, we demonstrate that a model based on a
Ginzburg-Landau type of approach provides an adequate description
of both the magnetization process and of the phase diagram. The
developed model not only accounts correctly for the observed
spin-reorientation transitions, but it predicts also their unusual
angular dependence.

\end{abstract}


%75.30.Kz Magnetic phase boundaries (including magnetic
%transitions, metamagnetism, etc.)
%75.50.Ee Antiferromagnetics


\maketitle

\section{Introduction}
Phase diagrams of antiferromagnets in externally applied magnetic
fields have been studied, both experimentally and theoretically,
for more than 50 years.\cite{landau, nagamiya} In conventional
collinear antiferromagnets the competition between Zeeman,
anisotropy, and exchange interactions gives rise to several phase
transitions. For example, if the antiferromagnetic order parameter
has a preferred direction (easy axis), a magnetic field applied
along that axis provokes a spin-flop transition at a critical
field value $H_c$,  where  the loss in anisotropy energy is
compensated by a gain in Zeeman energy. Another transition in
standard antiferromagnets is the spin-flip transition, which
occurs when the Zeeman energy exceeds the exchange energy.

In the more complicated cases of non-collinear and/or frustrated
antiferromagnets the choice of an ordered phase and the
orientation of the relevant order parameter are dictated by a fine
balance between the different interactions or by fluctuation
effects. \cite{gso,fluct} This close competition implies rich
phase diagrams with unusual features (such as, e.g., the
appearance of magnetization plateaus),
 which have been studied both theoretically\cite{fluct,honecker}
 and experimentally.\cite{plateau-spinel,plateau-cs2cubr4}

In view of the above, it came as a surprise when the
quasi-one-dimensional Heisenberg antiferromagnet \bacusio{}
(hereafter BCSO), identified at low fields as an easy-axis
collinear antiferromagnet, revealed ``extra'' spin-reorientation
transitions, both in an applied field along the easy
axis,\cite{2sf, zheludev-magnstruct} as well as in transverse
applied fields.\cite{ultrasound,glazkov-afmr} By now, the phase
diagram for applied fields along the main directions of the
orthorhombic crystal unit cell is well
established:\cite{glazkov-phasediagr} at 2 K, with the field
applied along the easy axis $c$, two spin-reorientations are
observed (at $H_{c1}=18.9$ kOe and $H_{c2}=47.1$ kOe), with the
field applied along the $b$ axis, one spin-reorientation is
observed at $H_{c3}=73.9$ kOe and, finally, with the field applied
along the $a$ direction another spin-reorientation transition
takes place at $H_{c4}=114$ kOe.

On the basis of neutron scattering
experiments,\cite{zheludev-magnstruct} the transitions occurring
in a magnetic field applied along the easy axis were interpreted
as consecutive rotations of the order parameter away from the easy
axis, to a plane normal to it, followed by a rotation within this
plane. A weak noncollinearity in the spin-flopped phase, at
$H_{c1}<H<H_{c2}$, was suggested as
well.\cite{zheludev-magnstruct} Subsequent analyses of magnetic
resonance data confirmed that transitions in the transverse
direction were indeed rotations of the sublattice magnetization
away from the easy axis.

The temperature dependence of the static magnetization (see e.g.\
Ref.~\onlinecite{2sf}) exhibits unusual features at low
temperatures. The results of neutron-scattering experiments
confirmed the one-dimensional character of the spin system of
BCSO\cite{kenzelman} and the magnetic susceptibilities along the
$a$, $b$ and $c$ direction indeed exhibit the Bonner-Fisher maxima
expected for spin chains at elevated temperatures.\cite{2sf} At
lower temperatures, however, unexpected increases of $\chi_b$ and
$\chi_c$ with decreasing temperature are observed above the N\'eel
temperature $T_{\mathrm{N}}$, atypical for this type of spin
systems. More recent experiments probing the ${}^{29}$Si NMR line
shift and its temperature dependence also revealed deviations from
the expected conventional Bonner-Fisher behaviour.\cite{NMR}

The theoretical description of the physics underlying these phase
transitions is neither complete nor satisfactory. A
phenomenological approach was used to describe the low-temperature
phase transitions and the antiferromagnetic resonance
spectra.\cite{glazkov-afmr, glazkov-phasediagr} The peculiarities
of the susceptibility above the N\'eel temperature were discussed
in relation with the known behavior of weak
ferromagnets.\cite{glazkov-kvn-arxiv} Yet, these approaches
predict an unusual increase of certain parameters with respect to
their conventional estimates, hence requiring an additional
refinement of the theory. Mean-field theory models have been
attempted in the past,\cite{sato} but they require the exact
knowledge of many (often unavailable) microscopic parameters. By
contrast, a thermodynamics-based approach has better chances to
capture the overall physical picture, while being less demanding
in terms of parameter knowledge.

Recent NMR studies\cite{NMR} have provided a direct access to the
local magnetization, both above and below the N\'eel temperature,
revealing that a field-induced transverse magnetization appears
on the magnetic ions. The related staggered magnetic field is an
additional parameter that needs to be considered in a
comprehensive discussion of the low-temperature magnetic features
of BCSO.

In this work, based on a conventional Ginzburg-Landau (GL)
approach which takes into account the new experimental findings,
we reconsider the interpretation of the phase transitions in BCSO.
The field-induced transverse staggered magnetization (TSM)
competes with the order parameter of the phase with spontaneously
broken symmetry below $T_{\mathrm{N}}$. This competition may cause
additional phase transitions in non-zero external magnetic fields
and hence influence the magnetization process. We demonstrate that
a GL-type analysis of the available data provides a
semi-quantitative description of the phase diagram and predicts
both the phase boundaries and their variation upon changing the
external magnetic-field orientation with respect to the crystal
axes.

In Sec.\ \ref{sec:theor_model} we describe and justify the
phenomenological model, whose parameters are obtained from fitting
the model calculations to experimental data. This is shown and
discussed in detail in Sec.\ \ref{sec:comparison}, leading finally
to some conclusions at the end of the manuscript.

\section{\label{sec:theor_model}The theoretical model}

We start our discussion from the paramagnetic phase. As the
temperature decreases towards the transition temperature, the
thermodynamic functions can be expanded over powers of the
different order parameters (i.e., over different irreducible
representations). The symmetry group of \bacusio{} ($Pnma$ or
$D_{2h}^{16}$, see Ref.~\onlinecite{structure}) includes only
one-dimensional representations which have been classified
earlier.\cite{glazkov-kvn-prb,glazkov-kvn-arxiv} Here, for the
sake of consistency, we will use the same classification and axes
notation ($x \parallel a$, $y \parallel b$, and $z \parallel c$).

\begin{figure}[tbh]
  \centering
  \includegraphics[width=\figwidth]{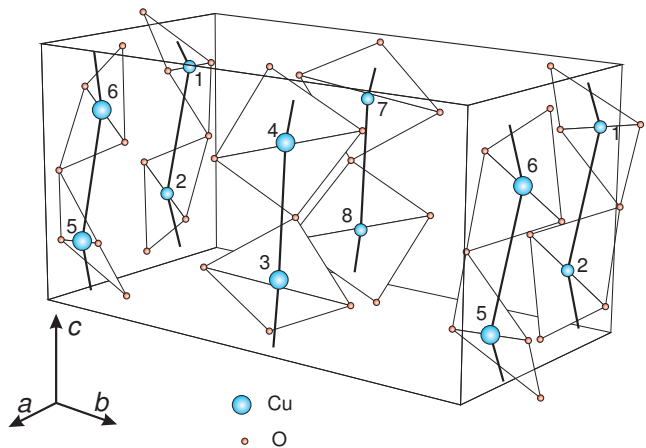}
  \caption{Positions of the magnetic ions and their oxigen surroundings in the unit cell of
  \bacusio{}. The fractional cell coordinates of the Cu$^{2+}$ ion (1)
  are given\cite{structure} by $(1/4 - 0.028, 0.004, 3/4 + 0.044)$. Chains direction is shown by bold line. }
\label{fig:struct}
\end{figure}

\begin{table}
\caption{Definitions of the relevant magnetic vectors and effect
of the symmetry operations on their components. Ions positions are
enumerated following Figure \ref{fig:struct}.  Signs in the first
column show  effect of the symmetry operations $I$
($(x,y,z)\rightarrow(-x,-y,-z)$),
   $C^2_z$ ($(x,y,z)\rightarrow(\frac{1}{2}-x,-y,\frac{1}{2}+z)$) and $C^2_y$
   ($(x,y,z)\rightarrow(-x,\frac{1}{2}+y,-z)$), correspondingly.
\label{tab:represent}}
\begin{ruledtabular}
\begin{tabular}{l}
    $\vect{L}_{1}=\vect{S}_{1}-\vect{S}_{2}-\vect{S}_{3}+\vect{S}_{4}+\vect{S}_{5}-\vect{S}_{6}-\vect{S}_{7}+\vect{S}_{8}$  \\
    $\vect{L}_{2}=\vect{S}_{1}-\vect{S}_{2}+\vect{S}_{3}-\vect{S}_{4}+\vect{S}_{5}-\vect{S}_{6}+\vect{S}_{7}-\vect{S}_{8}$  \\
    $\vect{L}_{4}=\vect{S}_{1}-\vect{S}_{2}-\vect{S}_{3}+\vect{S}_{4}-\vect{S}_{5}+\vect{S}_{6}+\vect{S}_{7}-\vect{S}_{8}$ \\
    $\vect{L}_{6}=\vect{S}_{1}-\vect{S}_{2}+\vect{S}_{3}-\vect{S}_{4}-\vect{S}_{5}+\vect{S}_{6}-\vect{S}_{7}+\vect{S}_{8}$ \\
\end{tabular}
\begin{tabular}{cc}
   $+++$ & $L_{1x}$, $L_{2y}$ \\
    $++-$ & $L_{2x}$, $L_{1y}$, $H_z$ \\
    $+-+$ & $H_y$, $L_{1z}$ \\
    $-++$ & $L_{4x}$, $L_{6y}$ \\
    $--+$ & $L_{4z}$ \\
    $-+-$ & $L_{6x}$, $L_{4y}$ \\
    $+--$ & $H_x$, $L_{2z}$ \\
    $---$ & $L_{6z}$ \\
\end{tabular}
\end{ruledtabular}
\end{table}

Because of the quasi-one dimensionality of \bacusio{}, its
magnetic order, involving spin-$\nicefrac{1}{2}$ Cu$^{2+}$ ions,
appears at temperatures much lower than those implied by the
in-chain exchange interaction strength ($T_{\mathrm{N}} \approx
9.2$~ K,\cite{2sf,glazkov-phasediagr} while
$J=24.1$~meV\cite{kenzelman}). Thus, at the phase transition,
representations corresponding to the ferromagnetic in-chain order
can be totally ruled out. Among the eight irreducible
representations, only four correspond to the antiferromagnetic
in-chain ordering. These differ by the mutual orientation of the
spins in the transverse direction and, following
Ref.~\onlinecite{glazkov-kvn-prb}, we denote them by $\vect{L}_1$,
$\vect{L}_2$, $\vect{L}_4$, and $\vect{L}_6$. These
representations and effect of symmetry operations on the
corresponding components of magnetic vectors are shown in the
Table \ref{tab:represent}. $\vect{L}_6$ represents the ordering
pattern in the form of a ferromagnetic alignment of neighbouring
spins along the $a$ axis and an antiferromagnetic (AFM) alignment
of neighbouring spins along the $b$ axis. $\vect{L}_1$ instead
represents an AFM alignment along both the $a$- and the $b$ axis.
An AFM alignment along $a$, but FM alignment along $b$ is
represented by $\vect{L}_2$. Finally, $\vect{L}_4$ represents an
FM alignment along both the $a$- and the $b$ axis. All of them are
consistent with an AFM order along the $c$ axis.

In the absence of an applied field the main order parameter
$\vect{L}_6$ develops at the N\'eel point, as established by
neutron scattering experiments.\cite{zheludev-magnstruct} Bilinear
invariants that couple different representations with the
principal order parameter and magnetic field can be directly
deduced from Table \ref{tab:represent}:
\begin{equation}\label{eqn:bilinear}
L_{6x}L_{4y},~L_{6y}L_{4x},~L_{1y}H_z,~L_{1z}H_y,~L_{2x}H_z,~L_{2z}H_x
\end{equation}

Thus, while the components of $\vect{L}_4$ can appear only
simultaneously with the corresponding components of the main order
parameter $\vect{L}_6$, and are exactly zero above
$T_{\mathrm{N}}$, the components of $\vect{L}_1$ and $\vect{L}_2$
can be induced by an applied magnetic field, independent of the
main-order parameter's existence or orientation. If $\vect{L}_1$
(or $\vect{L}_2$) would be the principal order parameter,
invariants $L_{1y}H_z$ and $L_{1z}H_y$ (or $L_{2x}H_z$ and
$L_{2z}H_x$) would lead to the appearance of the weak
ferromagnetism, as it happens in the related compound
BaCu$_2$Ge$_2$O$_7$.\cite{tsukada-bcgo}

The representations $\vect{L}_1$, $\vect{L}_2$, and $\vect{L}_4$
do not need to be included in the free-energy expansion if only
the macroscopic energy is of interest, minimization over their
components results  in the thermodynamic function expansion over
principal order parameter. At the microscopic level, however,
these representations are components of the local magnetization
which are experimentally accessible.

In particular, recent NMR studies\cite{NMR} have demonstrated the
presence of a field-induced staggered magnetization, both above
and below $T_{\mathrm{N}}$. A similar effect is well known in the
case of weak ferromagnets above the transition
temperature.\cite{borovikozhogin,borovik:lectures} In order to
capture this situation in our model, the field-induced transverse
staggered magnetization, here represented by  $\vect{L}_1$ and
$\vect{L}_2$, have to be included in the GL free-energy expansion.
For our purposes and in order to simplify the calculations, we
refrain from considering  $\vect{L}_4$, which would only lead to a
slight renormalization of the anisotropy constants.

The value of the critical exponent describing the growth of
$\vect{L}_6$ close to $T_{\mathrm{N}}$, as determined by neutron
scattering\cite{kenzelman} and by NMR\cite{NMR} is close to 0.25,
differing significantly from the classical GL value of 0.5.
Therefore, a GL approach in dealing with the present case has its
limitations. Yet, it can still provide useful insights into the
nature of the phase transitions and the understanding of competing
magnetic order parameters of quasi-one-dimensional spin systems.

Following the general theory\cite{landau} we use the thermodynamic
function $\widetilde\Phi$, defined in such a way that
$\partial{\widetilde\Phi}/\partial{\vect{H}}=-\vect{H}/(4\pi)-\vect{M}$:

\begin{eqnarray}
   \widetilde {\Phi}&=&\Phi_0+A_6 \vect{L}_6^2+A_1 \vect{L}_1^2+A_2 \vect{L}_2^2+\nonumber\\
   &&+B_6 \vect{L}_6^4+B_{16}\vect{L}_6^2\vect{L}_1^2+B_{26} \vect{L}_6^2\vect{L}_2^2+\nonumber\\
   &&+B'_{16} (\vect{L}_6\cdot\vect{L}_1)^2+B'_{26} (\vect{L}_6\cdot\vect{L}_2)^2+\nonumber\\
   &&+D(\vect{H}\cdot\vect{L}_6)^2+D'\vect{ H}^2\vect{L}_6^2+\nonumber\\
   &&+a_x L_{6x}^2+a_y L_{6y}^2+ \alpha_y L_{1z} H_y +\alpha_z L_{1y} H_z+\nonumber\\
   &&+\beta_z L_{2x}H_z+\beta_x L_{2z}H_x-\nonumber\\
   &&-{\frac{1}{2}}\chi_p\vect{H}^2-{\frac{1}{2}}\gamma_x H_x^2-{\frac{1}{2}}\gamma_y H_y^2-\frac{\vect{H}^2}{8\pi}.
\label{eqn:expansion}
\end{eqnarray}

\noindent The quadratic terms $A_i\vect{L}_i^2$ describe the
exchange rigidity towards the formation of the corresponding order
parameter. As usual, $A_6=\xi_6(T-T_{\mathrm{N}}^{(0)})$, while
$A_1$ and $A_2$ remain positive. We suppose that, because of the
low dimensionality of the spin system, $A_{1,2}$ are particulary
small close to $T_{\mathrm{N}}$. Their temperature dependence can
be relatively strong, however, and thus we assume a linear
temperature dependence in the vicinity of $T_{\mathrm{N}}$, i.e.:
$A_{1,2}(T)=A_{1,2}^{(0)}\,[1+\xi_{1,2}^{(rel)}(T-T_{\mathrm{N}}^{(0)})]$.
All other coefficients are postulated to be temperature
independent. The fourth-order term $B_6 \vect{L}_6^4$ fixes the
magnitude of the principal order parameter $\vect{L}_6$ below the
transition. The following terms $B_{i6}\vect{L}_6^2\vect{L}_i^2$
and $B'_{i6} (\vect{L}_6\cdot\vect{L}_i)^2$ describe the
competition between the field-driven TSM $\vect{L}_{1,2}$ and the
exchange-driven principal order parameter $\vect{L}_6$, the key
topic of our paper. The next terms describe the usual exchange
contributions to the magnetization ($D$ and $D'$), the anisotropic
interactions affecting the principal order parameters ($a_x$ and
$a_y$), and the anisotropic interactions responsible for the
coupling of $\vect{L}_{1,2}$ to the magnetic field ($\alpha_{y,z}$
and $\beta_{x,z}$). Finally, the last line represents the
paramagnetic susceptibility $\chi_p$ and its corrections (e.g.,
due to $g$-factor anisotropy) $\gamma_x$ and $\gamma_y$.

The magnetic anisotropy axes of \bacusio{} have been identified by
magnetization,\cite{2sf} neutron
scattering\cite{zheludev-magnstruct} and antiferromagnetic
resonance (AFMR)\cite{hayn-afmr,glazkov-afmr} experiments: the $c$
axis represents the easy axis, while $b$ is the secondary easy
axis (hence implying $a_x>a_y>0$). The $B_{i6}$ coefficients are
expected to be positive when non-coexisting representations are
favored. The $B'_{i6}$ coefficients instead, which are responsible
for the preferred mutual orientation of the principal order
parameter and TSM, do not have an a priori given sign.

The field-induced  TSMs $\vect{L}_{1,2}$ can be found by
minimizing the value of $\widetilde {\Phi}$, as given by
Eq.~(\ref{eqn:expansion}). For instance, when $\vect{H} \parallel
y$ the only non-zero component is
\begin{equation}\label{eqn:l1z}
    L_{1z}=-\frac{\alpha_yH_y}{2(A_1+B_{16}\vect{L}_6^2+B'_{16}L_{6z}^2)}.
\end{equation}
%
%\noindent
This  TSM can be induced by the magnetic field already above
$T_{\mathrm{N}}$. At the N\'eel point $\vect{L}_6$ starts to grow
and to suppress $L_{1z}$ ($A_1,B_{16}>0$). Similarly, for the
other two principal field orientations, $\vect{H} \parallel x$
induces $L_{2z}$, while $\vect{H} \parallel z$ induces both
$L_{1y}$ and $L_{2x}$. All the induced  transverse staggered
magnetizations depend linearly on the applied field.

The substitution of the field-induced TSMs found above into
Eq.~(\ref{eqn:expansion}) results in field-dependent terms
quadratic in $H$. These terms correspond to  the corrections to
the susceptibility which, in case of applied fields along the
crystalline axes, can be written as:
\begin{eqnarray}
  \Delta\chi_x&=&\frac{1}{2}\frac{\beta_x^2}{A_2+B_{26}\vect{L}_6^2+B'_{26}L_{6z}^2}\label{eqn:chi_x}\\
  \Delta\chi_y&=&\frac{1}{2}\frac{\alpha_y^2}{A_1+B_{16}\vect{L}_6^2+B'_{16}L_{6z}^2}\label{eqn:chi_y}\\
  \Delta\chi_z&=&\frac{1}{2}\frac{\alpha_z^2}{A_1+B_{16}\vect{L}_6^2+B'_{16}L_{6y}^2}+\nonumber\\
  &+&\frac{1}{2}\frac{\beta_z^2}{A_2+B_{26}\vect{L}_6^2+B'_{26}L_{6x}^2}\label{eqn:chi_z}
\end{eqnarray}

These corrections to the susceptibility provide a natural
explanation for the additional contributions to $\chi$ observed
above $T_{\mathrm{N}}$. Below $T_{\mathrm{N}}$ the $\Delta\chi$
terms depend on the orientation of the main order parameter and,
hence, provide clues about the possible spin-reorientation
transitions. For instance, for $\vect{H} \parallel x$, a transition
with the rotation of the order parameter from the easy-axis $z$
towards the $y$-axis is possible only if $B'_{26}>0$. The
corresponding transition field is
\begin{equation}\label{eqn:Hc4}
  H_{c4}^2=\frac{4 a_y A_2^2}{\beta_x^2 B'_{26}}.
\end{equation}
%
%\noindent
Note that the $a_y$ and $\beta_x^2$ parameters are of the same
order of magnitude as the corrections due to the spin-orbit
interaction. Consequently, in the ordinary antiferromagnet, the
field $H_{c4}$ should be comparable with the exchange field. The
relatively small (as compared with the exchange field defined by
the in-chain exchange integral) value of $H_{c4}$ is in fact due
to the tiny value of $A_2$, in turn related to the
one-dimensionality of the system. The positiveness of $B'_{26}$
means that at high applied fields an orthogonal alignment of the
field-induced TSM $\vect{L}_2$ and of the exchange-driven
$\vect{L}_6$ order parameters is favored, while at low fields (for
$\vect{H}
\parallel x$) both $\vect{L}_2$ and $\vect{L}_6$ are parallel to
$z$.
As a result, since the field-induced order parameter is determined
by the applied field, the main order parameter will start to rotate
whenever the susceptibility-related energy gains overcome the
anisotropy-related losses.

Similarly, for $\vect{H} \parallel y$, $B'_{16}>0$ will cause a
spin-reorientation at a critical field
\begin{equation}\label{eqn:Hc3}
  H_{c3}^2=\frac{4 a_x A_1^2}{\alpha_y^2 B'_{16}}.
\end{equation}
%
%\noindent
Finally, for $\vect{H} \parallel z$, a rotation of the main order
parameter from the secondary easy-axis $y$ towards the hard axis $x$
is possible if
\begin{equation}\label{eqn:Hc2-condition}
  \frac{\alpha_z^2B'_{16}}{A_1^2}-\frac{\beta_z^2B'_{26}}{A_2^2}>0,
\end{equation}
%\noindent
at an applied field
\begin{equation}\label{eqn:Hc2}
  H_{c2}^2=\frac{4(a_x-a_y) A_1^2 A_2^2}{ \alpha_z^2 B'_{16} A_2^2 - \beta_z^2 B'_{26} A_1^2}.
\end{equation}

The normal spin-flop occurs at the field:
\begin{equation}\label{eqn:Hc1}
  H_{c1}^2=\frac{a_y}{D - {\alpha_z^2 B'_{16}}/(4A_1^2)}.
\end{equation}

Likewise, field-induced shifts of the N\'eel temperature can be
straightforwardly calculated from Eq.~(\ref{eqn:expansion}). As an
example, we consider the main order parameter to be oriented as in
the case of \bacusio{} and obtain:\\ For $\vect{H} \parallel x$
and $H>H_{c4}$ (i.e., $\vect{L}_6 \parallel y$):
\begin{equation}\label{eqn:TN-x}
  T_{\mathrm{N}}=T_{\mathrm{N}}^{(0)}-\frac{1}{\xi_6}\left[a_y+\left(D'+\frac{\beta_x^2B_{26}}{4A_2^2}\right)
  H^2\right].
\end{equation}

\noindent
For $\vect{H} \parallel y$ and $H>H_{c3}$ (i.e., $\vect{L}_6 \parallel x$):
\begin{equation}\label{eqn:TN-y}
  T_{\mathrm{N}}=T_{\mathrm{N}}^{(0)}-\frac{1}{\xi_6}\left[a_x+\left(D'+\frac{\alpha_y^2B_{16}}{4A_1^2}\right)
  H^2\right].
\end{equation}

\noindent
For $\vect{H} \parallel z$ and $H>H_{c2}$ (i.e., $\vect{L}_6 \parallel x$):
\begin{eqnarray}\label{eqn:TN-z}
  T_{\mathrm{N}}&=&T_{\mathrm{N}}^{(0)}-\frac{1}{\xi_6}\Biglb[ a_x+\Biglb(D'+\frac{\alpha_z^2B_{16}}{4A_1^2}\nonumber\\
&&+\frac{\beta_z^2(B_{26}+B'_{26})}{4A_2^2}\Bigrb)H^2\Bigrb].\label{eqn:TN-z}
\end{eqnarray}

In case of \bacusio{} the N\'eel temperature was found to increase
when $\vect{H} \parallel x$ and to decrease for applied fields
along the other two directions. This means that additional
corrections turn out to be comparable in magnitude with the main
exchange term $D' \vect{H}^2$, which again can be explained by the
particular smallness of the $A_{1,2}$ parameters.

The paramagnetic-antiferromagnetic phase boundary was studied in
the mean-field approach considering coupled chains in a staggered
field.\cite{sato} This approach demonstrated the suppression of
the symmetry-breaking ordered phase by the staggered field. Our
results show a similar behavior: positive $B_{16}$ and $B_{26}$
(which corresponds to the competition between the principal order
parameter and the TSM) leads to the decrease of the N\'eel
temperature with respect to the ordinary case $B_{16,26}=0$.
Besides that, our results show that under specific conditions
(corresponding to the positiveness of $B'_{16,26}$ in our
thermodynamic model) the transverse staggered field affects not
only the stability of the ordered phase but necessarily leads to
new spin-reorientation transitions in the ordered phase.

\section{\label{sec:comparison}Comparison of the model with experimental data and discussion} %

To compare the theoretical model with experimental data we will rely
mostly on the already published phase diagram\cite{glazkov-phasediagr}
and on NMR and magnetization data.\cite{NMR}

The expansion (\ref{eqn:expansion}) of the thermodynamic function
$\widetilde {\Phi}$ includes 19 explicit parameters and 2
coefficients $\xi_{1,2}^{(rel)}$, which describe the temperature
dependence of $A_{1,2}$. Although the existing experimental data
allow us to fix all the parameters, part of them, however, are not
critical for the computation of the phase diagram. Since our
procedure is affected by the choice of certain extrapolations (see
below), the values used here differ slightly from those reported
in Ref.~\onlinecite{NMR}.

Since all the measurable parameters, except the absolute
magnitudes of the field-induced order parameters, depend only on
the ratios $\alpha_{y,z}^2/A_1$, $\beta_{x,z}^2/A_2$,
$B_{16}/A_1$, $B'_{16}/A_1$, $B_{26}/A_2$, and $B'_{26}/A_2$, the
values of the $A_{1,2}$ terms at $T_{\mathrm{N}}$
($A_{1,2}^{(0)}$) can be evaluated independently.

\begin{figure}[tbh]
  \centering
  \includegraphics[width=\figwidth]{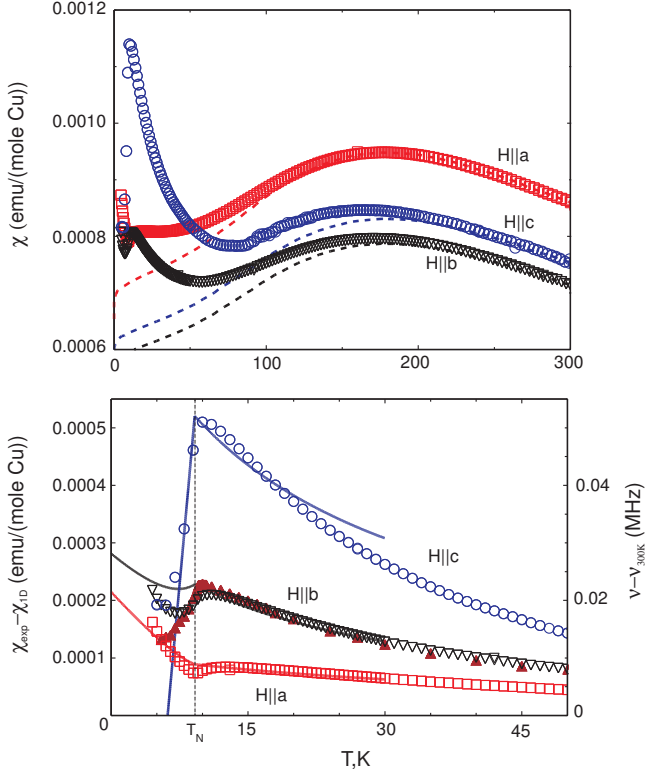}
  \caption{(color online) Upper panel: Temperature
  dependence of the magnetic susceptibilities measured at 10 kOe
  (Ref.~\onlinecite{NMR}) (symbols). Model curves for a pure 1D
  Heisenberg antiferromagnet (Ref.~\onlinecite{klumper}) with
  $J=24.1$~meV, $g_a=2.20$, $g_b=2.00$, and $g_c=2.06$ (dashed lines).
  Lower panel: Difference between the measured $\chi(T)$ and the
  corresponding model curves (open symbols), variation of the
  NMR shift\cite{NMR} with temperature
  (below $T_{\mathrm{N}}$ the average position of the split peaks is shown)
  (closed symbols), model calculations of the present work (solid lines).}\label{fig:magn}
\end{figure}

By requiring that the saturated value of the main order parameter
at zero applied field is unity and by using the known magnitude of
the zero-field specific heat jump,\cite{glazkov-phasediagr}
$\Delta C=0.61$ J/(K$\cdot$mol) (per mole of compound), we obtain
$\xi_6=0.305\times 10^7$ emu/(K$\cdot$ mol Cu) and
$B_6=1.40\times10^7$ emu/(mol Cu). Paramagnetic contributions to
the susceptibility can be determined by an extrapolation of the
susceptibility curve for 1D Heisenberg chains.\cite{klumper} To
this end we used the exchange integral value $J=24.1$~meV,
determined by neutron scattering experiments,\cite{kenzelman} and
fitted the high-temperature tails of the measured $\chi(T)$ curves
by setting the $g$-factor values equal to 2.20, 2.00 and 2.06 for
$\vect{H}
\parallel a$, $b$ and $c$, respectively (see Fig.~\ref{fig:magn}).
This extrapolation at the N\'eel temperature yields
$\chi_p=6.29\times 10^{-4}$ emu/(mol Cu), $\gamma_x=0.88\times
10^{-4}$ emu/(mol Cu), $\gamma_y=-0.34\times 10^{-4}$ emu/(mol
Cu).

A subtraction of the corresponding Bonner-Fisher curve from the
measured magnetization data (see Fig.~\ref{fig:magn}) provides the
additional contributions to the susceptibility, as described by
Eqs.~(\ref{eqn:chi_x})--(\ref{eqn:chi_z}). By extrapolating the
latter curves to the $T_{\mathrm{N}}$ value, we can fix the
following parameters combinations:
$\Delta\chi_x(T_{\mathrm{N}})=\frac{\beta_x^2}{2A_2}=0.89\times10^{-4}$
emu/(mol Cu),
$\Delta\chi_y(T_{\mathrm{N}})=\frac{\alpha_y^2}{2A_1}=2.28\times10^{-4}$
emu/(mol Cu) and
$\Delta\chi_z(T_{\mathrm{N}})=\frac{\alpha_z^2}{2A_1}+\frac{\beta_z^2}{2A_2}=5.20\times10^{-4}$
emu/(mol Cu).
The coefficients $\xi_{1,2}^{(rel)}$ can be estimated from the
temperature dependence of the NMR shifts across the transition and
from the additional contributions to the susceptibility above
$T_{\mathrm{N}}$: $\xi_1^{(rel)}=0.037$ K$^{-1}$ and
$\xi_2^{(rel)}=0.02$ K$^{-1}$. The ratio of the anisotropy
constants $a_x/a_y=3.4$ is known from AFMR
data.\cite{glazkov-afmr} Finally, the main correction to the
susceptibility, due to the onset of an antiferromagnetic order,
$D$, can be estimated from the magnetization curves as
$D=8.0\times10^{-4}$ emu/(mol Cu).

\begin{table}
\caption{Parameters of the thermodynamic function $\widetilde
{\Phi}$, as given by Eq.~(\ref{eqn:expansion}), used in the model
calculations.\label{tab:parameters}}
\begin{ruledtabular}
\begin{tabular}{cc}
    $\chi_p$&$6.29\times 10^{-4}$ emu/(mol Cu)\\
    $\gamma_x$&$0.88\times 10^{-4}$ emu/(mol Cu)\\
    $\gamma_y$&$-0.34\times 10^{-4}$ emu/(mol Cu)\\
    $D$&$8.0\times10^{-4}$ emu/(mol Cu)\\
    $D'$&$-0.88\times10^{-4}$ emu/(mol Cu)\\
    $\xi_1^{(rel)}$&$0.037$ \text{K}$^{-1}$\\
    $\xi_2^{(rel)}$&$0.02$ \text{K}$^{-1}$\\
    $\xi_6$& $0.305\times10^7$ emu/(K$\cdot$ mol Cu)\\
    $B_6$& $1.40\times10^7$ emu/(mol Cu)\\
    $\beta_x^2/A_2^{(0)}$&$1.78\times10^{-4}$ emu/(mol Cu)\\
    $\alpha_y^2/A_1^{(0)}$&$4.56\times10^{-4}$ emu/(mol Cu)\\
    $\beta_z^2/A_2^{(0)}$&$2.03\times 10^{-4}$ emu/(mol Cu)\\
    $\alpha_z^2/A_1^{(0)}$&$8.37\times10^{-4}$ emu/(mol Cu)\\
    $B_{16}/A_1^{(0)}$&0.622\\
    $B_{26}/A_2^{(0)}$&1.197\\
    $B'_{16}/A_1^{(0)}$&0.892\\
    $B'_{26}/A_2^{(0)}$&0.332\\
    $a_x$&$6.76\times10^5$ emu/(mol Cu)\\
    $a_y$&$1.99\times10^5$ emu/(mol Cu)\\
    $A_1^{(0)}$&$1.06\times10^7$ emu/(mol Cu)\\
    $A_2^{(0)}$&$6.63\times10^5$ emu/(mol Cu)\\
\end{tabular}
\end{ruledtabular}
\end{table}

The few remaining parameters can be tuned to the extent that the
calculated critical-field values and the field-induced shifts of
$T_{\mathrm{N}}$ agree best with those obtained from the
experiment. The critical fields at the paramagnet-antiferromagnet
phase boundary  are $H_{c1}=18$~kOe, $H_{c2}=53$~kOe,
$H_{c3}=81.5$~kOe, and $H_{c4}=116$~kOe, while the values of the
N\'eel temperature shifts at 14~T ($\Delta T_{\mathrm{N}} =
T_{\mathrm{N}}^{(14\,\mathrm{T})}-T_{\mathrm{N}}^{(0)}$ are %equal to
$0.16(2)$~K for $\vect{H} \parallel x$, $-0.11(4)$~K for $\vect{H}
\parallel y$, and $-0.98(2)$~K for $\vect{H} \parallel
z$).\cite{glazkov-phasediagr} All the parameter values and their
relevant combinations are listed in Table~\ref{tab:parameters}.

\begin{figure}[t]
  \centering
  \includegraphics[width=\figwidth]{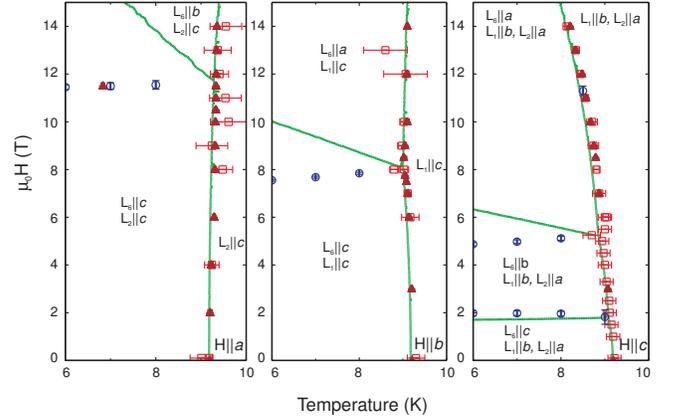}
  \caption{(color online) Experimental phase diagram of \bacusio{}
  (Ref.~\onlinecite{glazkov-phasediagr}) (magnetization data ---
  open circles and squares, specific heat measurements --- closed triangles) vs.\ modelled
  phase boundaries (solid lines). Orientations of the principal
  order parameter $\vect{L}_6$ and of the field induced TSM $\vect{L}_{1,2}$ are
  shown for the corresponding phases. }\label{fig:t-h}
\end{figure}

\begin{figure}[t]
  \centering
  \includegraphics[width=\figwidth]{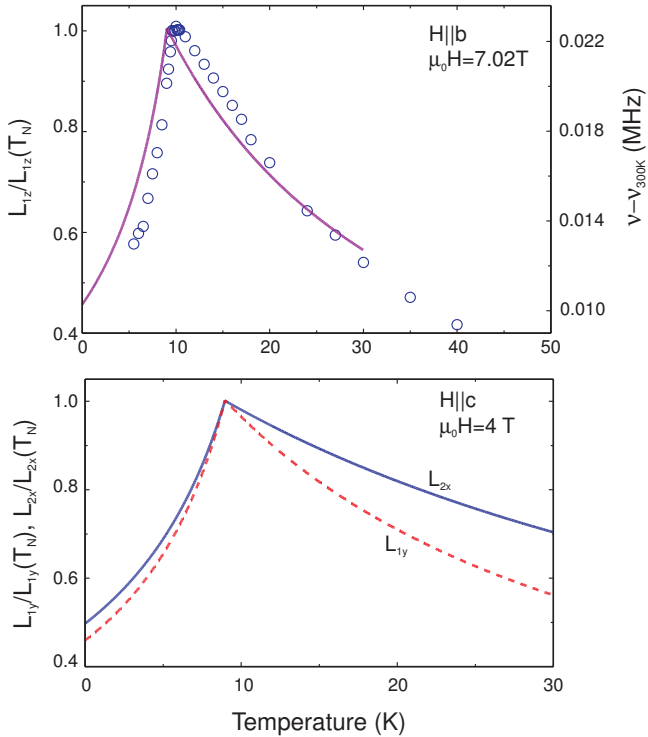}
  \caption{(color online) Upper panel: Comparison of the modelled
  temperature dependence of the field-induced
  order parameter (solid line) vs.\ measured
  NMR line shift\cite{NMR} ($\mu_0 H=7.02$~T, $\vect{H} \parallel b$) (circles).
  Lower panel: Temperature dependence of the induced order parameters
  for $\vect{H} \parallel c$ and $\mu_0H=4.0$~T.}\label{fig:l12}
\end{figure}

With these parameters the magnetization curves can be calculated
and the relevant phase boundaries can be established.
Figures~\ref{fig:magn} and \ref{fig:t-h} show that the modelled
curves are reasonably close to experiment. The main failure of
the model consists in the predicted temperature dependence of the
high-field spin-reorientation transitions, which is not observed
in the experiment. Probably this can be accounted for by
considering the role of thermal fluctuations, which are known to
stabilize collinear magnetic structures\cite{fluct} (field-induced
TSM and exchange-driven order parameter are collinear in the
low-field phases for $\vect{H}
\parallel a,b$). Our model provides an adequate
description also for the observed NMR shifts, both above and below
the N\'eel point, which, for $\vect{H} \parallel b$,\cite{NMR} is
mostly due to the staggered magnetization pattern related to
$L_{1z}$ (see Fig.~\ref{fig:l12}). The scaling of the NMR
shift\cite{NMR} and of $\Delta\chi_y$ above $T_{\mathrm{N}}$ (see
Fig.~\ref{fig:magn}) confirms once more the reliability of the model
[compare Eqs.~(\ref{eqn:l1z}) and (\ref{eqn:chi_y})].

The evaluation of the individual values of $A_{1,2}^{(0)}$
mentioned above is possible if the values of the field-induced
TSM's at the phase transitions in differently oriented magnetic
fields are known. The evaluation of the latter is not
straightforward but still feasible.

An estimate for the field-induced TSM, $L_{1z}
=0.035$~$\mu_{\mathrm{B}}$ per Cu$^{2+}$ ion (in an applied field
of 7.02~T), was obtained in recent NMR work.\cite{NMR} Since in
our model the saturated value of the main order parameter is
normalized, while its measured value\cite{zheludev-magnstruct} is
ca.\ $0.15~\mu_{\mathrm{B}}$ per Cu$^{2+}$ ion, this corresponds
to $L_{1z}(T_{\mathrm{N}})=0.23$ in normalized units. From the
latter value and from Eq.~(\ref{eqn:l1z}) we find
$A_1^{(0)}=1.06\times10^7$ emu/(mol Cu).

The $A_2^{(0)}$ value can be estimated from the canting of the
magnetic structure, as observed in neutron scattering
experiments\cite{zheludev-magnstruct} with $\vect{H} \parallel c$.
These experiments revealed an $L_{2x}$ component of ca.\ 0.17
(corresponding to a canting angle of $\simeq 10^{\circ}$ at $\mu_0
H \sim4$~T). However, because of the coincidence of the structural
and magnetic Bragg peaks in \bacusio{}, the determination of the
magnetic scattering intensities required the subtraction of the
peak intensities recorded just above $T_{\mathrm{N}}$. Since the
contribution of the field-induced order parameter is subtracted as
well, the observed low-temperature magnitude of $L_{2x}$
represents, in fact, only the change of $L_{2x}$ across
$T_{\mathrm{N}}$ due to the competition between the field-induced
and exchange-driven orders. Our calculations (Fig.~\ref{fig:l12})
show that $L_{2x}$ at 6~K amounts to $\sim 50\%$ of its value at
$T_{\mathrm{N}}$. Therefore, by assuming
$L_{2x}(T_{\mathrm{N}})=0.35$, we obtain
$A_2^{(0)}=6.63\times10^5$ emu/(mol Cu). Note that an
$L_{1y}$ component should also exist for $\vect{H} \parallel c$.
However, its magnitude at $T_{\mathrm{N}}$ and $\mu_0H=4$ T is, as
calculated using the found parameters, ca.\ 0.17 (i.e.\ only half
of $L_{2x}$). Besides, our model calculations show that $L_{1y}$
strongly changes below T$_N$ only at $H_{c1}<H<H_{c2}$ when the
principal order parameter is also aligned along the $y$-axis.
These reasons probably explain why $L_{1y}$ has not been observed
experimentally.
The above estimates for $A_{1,2}^{(0)}$ are consistent with their
expected small values. In fact, both of them are comparable with
$\xi_6(T-T_{\mathrm{N}})$ evaluated at $(T-T_{\mathrm{N}})\sim 1$ K.

\begin{figure}[t]
  \centering
  \includegraphics[width=\figwidth]{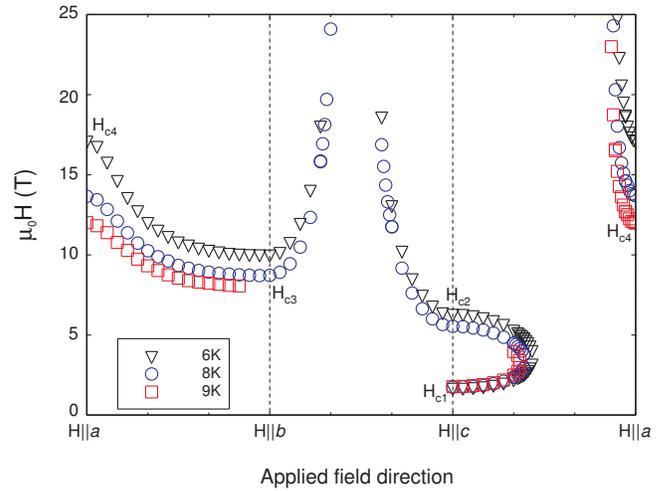}
  \caption{(color online) Modelled angular dependence of the spin-reorientation
  critical fields,
  performed at three different temperatures.
  The missing data for $T=9$~K in the rotation from
  $b$- toward the $c$-axis reflect the crossing of the PM-AFM boundary
  before a spin-reorientation could occur.}\label{fig:angular}
\end{figure}

Finally, we also computed the angular dependence of the critical
fields for spin-reorientation transitions (see
Fig.~\ref{fig:angular}). All the marked fields correspond to real
transitions, accompanied by a jump in the susceptibility and by a
sudden reorientation of the main order parameter $\vect{L}_6$. To
ensure that the angular dependence is not affected by changes in
magnitude of the main order parameter $\vect{L}_6$ (due to
closeness to the PM-AFM phase boundary), the modelling was carried
out at different temperatures (6, 8, and 9 K). All curves are
qualitatively similar, with differences mostly due to the above
mentioned temperature dependence of the upper critical fields
$H_{c2,3,4}$ and, partially (for the 9 K curve), to the crossing
of the PM-AFM phase boundary before a spin-reorientation
transition has occurred. Our model predicts quite a remarkable
angular dependence for the critical fields. On rotating from the
hard axis $a$ towards the secondary easy-axis $b$, the critical
field $H_{c4}$ transforms smoothly into $H_{c3}$. Upon further
rotation towards the easy axis $c$, the field required for a
spin-reorientation transition first grows very rapidly and, at an
intermediate critical angle, diverges to infinity. Subsequently,
it reappears from the high-field zone and finally converges to the
critical field $H_{c2}$. A rotation from the easy axis $c$ towards
the hard axis $a$ (right panel in Fig.~\ref{fig:angular})
demonstrates the merging of the two critical fields $H_{c1}$ and
$H_{c2}$ at a certain angle. Then the spin-reorientation transition
reappears from high fields and, finally, close to the $a$ axis,
evolves towards $H_{c4}$. To the best of our knowledge the angular
dependence of the critical fields in \bacusio{} has not yet been
studied. Its experimental investigation would represent an
independent additional test of the proposed model.

\section{Conclusions}
By making use of the Ginzburg-Landau theory of phase transitions
we propose a semi-quantitative description of the magnetic phase
diagram of 1D systems with competing interactions. We have
demonstrated that the competition between the field-induced and
the exchange-driven order parameters in a quasi-one-dimensional
antiferromagnet can lead to an unusual phase diagram and to
remarkable deviations of the magnetization process from that
expected in a 1D Heisenberg antiferromagnet. Additionally, in the
\bacusio{} model system, we predict an unusual angular dependence
of the critical fields, which will be object of future
experimental investigations.

\acknowledgments

V.G.\ thanks M.\ Zhitomirsky for the enlightening comments and
discussions. The present work was financially supported in part by
Russian Foundation for Basic Research and in part by the
Schweizerische Nationalfonds zur F\"{o}rderung der
Wissenschaftlichen Forschung (SNF) and the NCCR research pool
MaNEP of SNF.


\begin{thebibliography}{11}
%

\bibitem{landau}
L.~D.~Landau and E.~M.~Lifshitz, \textit{Electrodynamics of
Continuous Media}, 2nd ed., Course of Theoretical Physics, Vol.~8
(Pergamon Press, Oxford, 1984) Chap.~5.

\bibitem{nagamiya} T.~Nagamiya, K.~Yosida, and R.~Kubo, Adv.\ Phys. {\bf 4}, 1 (1955). % No. 13

\bibitem{gso}A.~S.~Wills, M.~E.~Zhitomirsky, B.~Canals, J.~P.~Sanchez, P.~Bonville, P.~Dalmas de R\'eotier and A.~Yaouanc
J.\ Phys.: Condens.\ Matter \textbf{18} L37 (2006)

\bibitem{fluct} A.~V.~Chubukov and D.~I.~Golosov,
J.\ Phys.: Condens.\ Matter \textbf{3}, 69 (1991).

\bibitem{honecker} A.~Honecker, J.~Schulenburg and J.~Richter,
J.\ Phys.: Condens.\ Matter \textbf{16} S749 (2004)


\bibitem{plateau-spinel} H.~Ueda, H.~Mitamura, T.~Goto, and Y.~Ueda,
Physical  Review B \textbf{73}, 094415 (2006)

\bibitem{plateau-cs2cubr4}
T.~Ono, H.~Tanaka, H.~Aruga Katori, F.~Ishikawa, H.~Mitamura, and
T.~Goto, Physical Review B \textbf{67}, 104431 (2003)

\bibitem{2sf} I.~Tsukada, J.~Takeya, T.~Masuda, and K.~Uchinokura,
Phys.\ Rev.\ Lett.\ \textbf{87}, 127203 (2001).

\bibitem{zheludev-magnstruct} A.~Zheludev, E.~Ressouche,
I.~Tsukada, T.~Masuda, and K.~Uchinokura, Phys.~Rev.~B {\bf 65},
174416 (2002).

\bibitem{ultrasound} M.~Poirier, M.~Castonguay, A.~Revcolevschi,
and G.~Dhalenne, Phys.\ Rev.\ B \textbf{66}, 054402 (2002).

\bibitem{glazkov-afmr} V.~N.~Glazkov, A.~I.~Smirnov, A.~Revcolevschi,
and G.~Dhalenne, Phys.\ Rev.\ B \textbf{72}, 104401 (2005).

\bibitem{glazkov-phasediagr} V.~N.~Glazkov, G.~Dhalenne,
A.~Revcolevschi, and A.~Zheludev, J. Phys.: Condens. Matter
\textbf{23}, 086003 (2011).




\bibitem{kenzelman} M.~Kenzelmann, A.~Zheludev, S.~Raymond,
E.~Ressouche, T.~Masuda, P.~B\"oni, K.~Kakurai, I.~Tsukada,
K.~Uchinokura, and R.~Coldea, Phys. Rev. B \textbf{64}, 054422
(2001).

\bibitem{NMR} F.~Casola, T.~Shiroka, V.~Glazkov, A.~Feiguin,
G.~Dhalenne, A.~ Revcolevschi, A.~Zheludev, H.-R.~Ott, and
J.~Mesot, arXiv:1207.1073, subm.\ to Phys.\ Rev.\ B.

%\bibitem{tokiwa} Y.~Tokiwa, T.~Radu, R.~Coldea, H.~Wilhelm, Z.~Tylczynski, and F.~Steglich,
%Phys.\ Rev.\ B \textbf{73}, 134414 (2006).

\bibitem{glazkov-kvn-arxiv}  V.~N.~Glazkov and H.-A.Krug von Nidda, arXiv:cond-mat/0210670.

\bibitem{sato}
%Coupled S=1/2 Heisenberg antiferromagnetic chains in an effective staggered field
M.~Sato and M.~Oshikawa, Phys.\ Rev.\ B {\bf 69}, 054406 (2004).




\bibitem{structure} J.~A.~S.~Oliveira, Ph.D.\ thesis,
Ruprechts-Karl Universit\"{a}t, Heidelberg, 1993.

\bibitem{glazkov-kvn-prb} V.~N.~Glazkov and H.-A.~Krug von Nidda,
Phys.\ Rev.\ B \textbf{69}, 212405 (2004).

\bibitem{tsukada-bcgo} I.~Tsukada, J.~Takeya, T.~Masuda, and K.~Uchinokura
Phys.\ Rev.\ B \textbf{62}, R6061 (2000).

\bibitem{borovikozhogin} A.~S.~Borovik-Romanov and V.~I.~Ozhogin,
Zh.\ Exp.\ Teor.\ Fiz.\ {\bf 39}, 27 (1960)
[Sov.~Phys.~JETP {\bf 12}, 18 (1961)].

\bibitem{borovik:lectures} A.~S.~Borovik-Romanov, \textit{Lectures on
Low-Temperature Magnetism}, (Novosibirsk University Press,
Novosibirsk, 1976) (in Russian).


\bibitem{hayn-afmr}  R.~Hayn, V.~A.~Pashchenko, A.~Stepanov, T.~Masuda, and K.~Uchinokura,
Phys.~Rev.~B {\bf 66}, 184414 (2002).

%\bibitem{bonner-fischer} J.~C.~Bonner and M.~E.~Fisher,
%Phys. Rev. \textbf{135}, A640 (1964).

\bibitem{klumper} A. Kl\"{u}mper and D. C. Johnston,
Phys.\ Rev.\ Lett.\ \textbf{84}, 4701 (2000).


\end{thebibliography}
\end{document}